\documentclass[final,3p,times,twocolumn]{elsarticle}

\usepackage{amssymb}
\usepackage{amsmath}
\usepackage{url}
\usepackage{csquotes}
\usepackage{bibunits}
\usepackage{siunitx}
\usepackage{xcolor}

\journal{Scripta}

\begin{document}

\begin{frontmatter}



\title{Unexpected Planar Dislocation Boundary Formation in FCC Metals Captured by Dark-Field X-ray Microscopy and Continuum Dislocation Dynamics}



\author[a]{Adam André William Cretton}
\author[b]{Khaled SharafEldin}
\author[c]{Axel Henningsson}
\author[c]{Felix Frankus}
\author[d]{Can Yıldırım}
\author[d]{Carsten Detlefs}
\author[d]{Flemming Bjerg Grumsen}
\author[e]{Albert Zelenika}
\author[b]{Anter El-Azab}
\author[c]{Grethe Winther}
\author[a]{Henning Friis Poulsen}

\affiliation[a]{organization={Department of Physics, Technical University of Denmark},
            city={Kongens Lyngby},
            postcode={2800}, 
            country={Denmark}}

\affiliation[b]{organization={School of Nuclear Engineering and School of Materials Engineering, Purdue University},
            city={West Lafayette},
            postcode={47907},
            country={USA}}

\affiliation[c]{organization={Department of Civil and Mechanical Engineering, Technical University of Denmark},
            city={Kongens Lyngby},
            postcode={2800},
            country={Denmark}}

\affiliation[d]{organization={European Synchrotron Radiation Facility (ESRF)},
            city={Grenoble},
            postcode={38000},
            country={France}}

\affiliation[e]{organization={IAM-MMI Mechanics of Materials 1 (WM1), Karlsruhe Institute of Technology (KIT)},
            city={Karlsruhe},
            postcode={76131},
            country={Germany}}

\begin{abstract}
Validating dislocation patterning models against \emph{in situ} imaging experiments is a longstanding goal in materials physics. Here, we provide the first direct morphological comparison of such models. Using \emph{in situ} Dark-Field X-ray Microscopy (DFXM), we map the local orientations in high-purity aluminium deformed along $[100]$ and find unexpected planar dislocation boundaries aligned with $\{111\}$ slip planes that form prior to the development of a conventional  dislocation cell structure. To explain this behaviour, we generate synthetic DFXM contrast images from a continuum dislocation dynamics (CDD) simulation. This mesoscale model, using nickel as a high stacking fault energy (SFE) FCC analogue, independently predicts the formation of the same $\{111\}$ planar boundary types. This correspondence demonstrates that state-of-the-art CDD and DFXM experimental data can be used synergistically  --- despite differences in strain rates and length scales --- as a practical route for refining continuum theories of plasticity.
\end{abstract}




\end{frontmatter}

Plastic deformation remains an open problem in metal physics, often likened in complexity to turbulence \cite{Cottrell2002}.  To date, no bottom-up theoretical framework can predict which dislocation structures form and how collective dislocation dynamics gives rise to the macroscopic mechanical response of crystalline materials. Current modelling approaches span multiple length scales. Discrete dislocation dynamics (DDD) explicitly resolves the motion and interaction of individual dislocations but is limited to small volumes and small applied strains \cite{madec2002, Cai2004, Frankus2025}. Continuum dislocation dynamics (CDD) treat dislocations and their interactions in a density-based framework, enabling mesoscale descriptions of the evolving dislocation microstructure \cite{Xia2016, Hochrainer2014, Sandfeld_2015}. Crystal plasticity models, by contrast,  capture anisotropic slip behaviour at the grain scale relying on phenomenological constitutive laws \cite{Ma2006, Roters2010}.

Among these approaches, CDD provides a natural framework for describing collective dislocation behaviour. Early CDD formulations laid important groundwork but operated within simplified interaction physics, small simulation volumes and restricted strain ranges, and thus could not reproduce the  experimentally observed spontaneous formation of dislocation structures. Recent theoretical and computational advances in CDD have overcome some of these barriers by improving the representation of cross slip, junction formation and annihilation processes within the CDD framework \cite{Lin_2020, VIVEKANANDAN2021104327}, allowing also for the simulation of volumes on the order of 10$\times$10$\times$10 $\mu$m$^3$ and applied strain up to $\sim$3-4\%,   approaching the limit of small-strain formulations. Finite-strain formalisms of CDD have also become available \cite{STARKEY2020103926}.

In parallel, experimental access to the early stages of dislocation patterning under bulk deformation has remained limited. Transmission electron microscopy (TEM) provides excellent spatial resolution but is inherently restricted to thin foils, small volumes, and predominantly post-mortem observations. Dark-Field X-ray Microscopy (DFXM) overcomes these limitations by enabling non-destructive, \emph{in situ} mapping of lattice orientation and elastic strain in bulk crystals with sub-micrometre spatial and milliradian angular resolutions \cite{Simons2015Dark-fieldCharacterization, Poulsen2017X-rayOptics}. Recent DFXM studies have directly visualised the emergence, evolution, and sharpening of dislocation cells and geometrically necessary boundaries in face centered cubic (FCC) metals in the 1\%- 5\% deformation range \cite{Zelenika20243DDeformation, Zelenika2025ObservingDeformation, Cretton2025MRL}. Furthermore, the development of a full-tensor reconstruction framework \cite{Henningsson2025, Detlefs2025, Kanesalingam2025} now provides a theoretical framework going towards quantitative comparison between experimentally measured and simulated microstructures.

In FCC metals with high stacking-fault energy, such as aluminium and nickel, plastic deformation proceeds by glide on multiple $\{111\}\langle110\rangle$ slip systems leading to the self-organisation of dislocations into characteristic substructures \cite{Holt1970DislocationMetals,Kocks2003}. TEM experiments have shown that the morphology of these substructures depends on the crystallographic orientation and the relative activity of the available slip systems. For grains with the tensile axis oriented along [100], the prevailing experimental picture is that deformation at intermediate strains (>5\%) gives rise to dislocation cells separated by incidental dislocation boundaries. In contrast, more planar, geometrically necessary boundaries are found in grains of all other orientations.  Boundaries near $\{111\}$  slip planes are associated with coplanar or near-coplanar slip conditions \cite{Hughes2003GNB,Huang2007DislocationDependence,Winther2007DislocationDependence}. Under uniaxial loading along the [100] direction, eight $\{111\}\langle110\rangle$ slip systems are equivalently oriented. In this high-symmetry case, the prevailing understanding is that deformation proceeds directly toward a cellular structure. 

Here we exploit the availability of \emph{in situ} bulk DFXM measurements together with the most recent CDD simulation tools to perform the first experimental test of the DFXM-CDD interface.  As a model system, we consider high stacking fault energy (SFE) FCC metals subjected to uniaxial [100] tensile deformation. We compare  the microstructure evolution in a pure aluminium single crystal measured  by DFXM with that predicted by CDD simulations of a pure Ni single crystal. To enable a rigorous and direct comparison, the simulated dislocation fields are propagated through a DFXM forward model to generate synthetic DFXM images. This approach places simulations and experiments on the same observational footing, allowing direct comparison between measured and predicted diffraction signatures. 

\begin{figure*}[!h]
    \centering
    \includegraphics[width=0.95\textwidth]{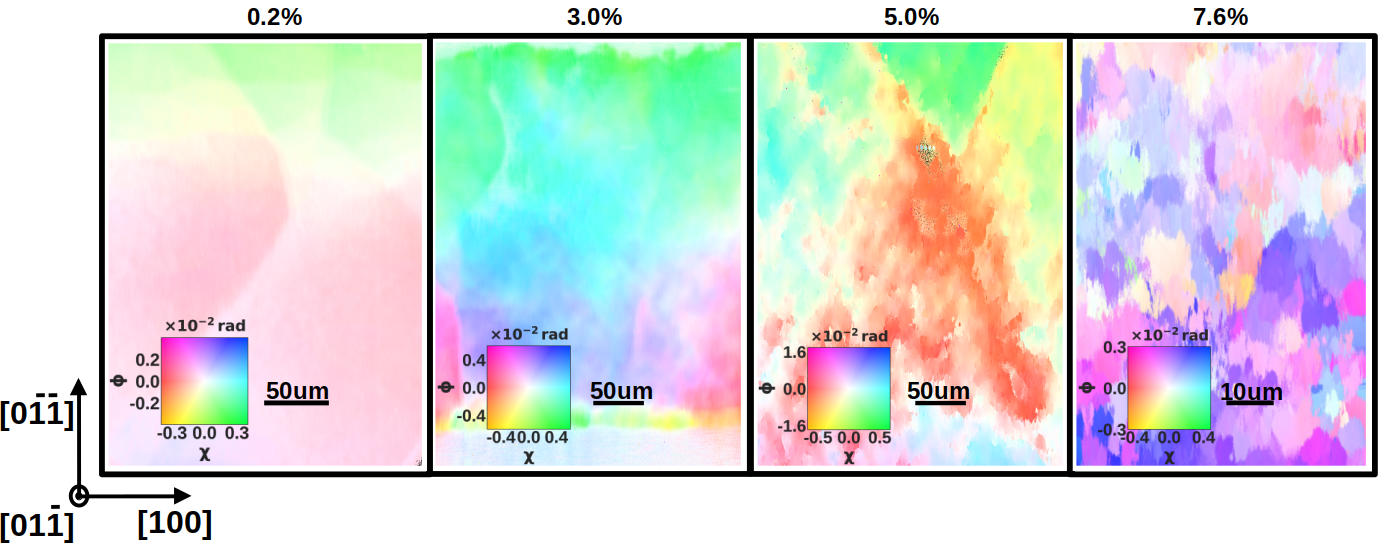}
    \caption{Summary of DFXM experiment. Evolution of lattice orientation during tensile deformation of Aluminium along the [100] axis is shown, visualised as orientation maps at 0.2\%, 3.0\%, 5.0\% and 7.6\% strain. Colour encodes the local lattice orientation. The illuminated layer was not tracked continuously across strain steps due to sample elongation.}
    \label{fig:mosaicity_maps}
\end{figure*}

 For the experiment, a rectangular specimen with nominal gauge dimensions of \(1 \times 1 \times 10\,\mathrm{mm}^3\) was cut out of a 99.9999\% pure single crystal using electrical discharge machining (EDM) to minimise surface damage. The sample was then annealed at \(540\,^{\circ}\mathrm{C}\) for 10 hours and furnace cooled to relieve residual stresses and homogenise the microstructure prior to deformation.

The experiment was conducted at the dedicated DFXM microscope at the ID03 beamline of the European Synchrotron Radiation Facility (ESRF) \cite{Isern2025} using a monochromatic 19 keV X-ray beam.  The details of the X-ray setup and microscope parameters are presented in Appendix A. Experimental data presented in this paper are available from the corresponding author upon reasonable request. Uniaxial tensile loading was applied \emph{in situ} using a custom-built load frame at a strain rate of approximately 0.001 s$^{-1}$ \cite{frankus2025thesis}. The macroscopic strain was measured using a single-camera digital image correlation (DIC) system; the applied stress was derived from load cell data. Stress-strain data are provided in Appendix D.

At discrete strains steps between 0.02\% and 7.6\% total elongation, the external load  was fixed. At each strain step DFXM images were acquired while scanning the sample in two orthogonal tilt directions $\phi$ and $\chi$, known as the rocking and rolling directions, see \cite{Poulsen2021Geometrical-opticsMicroscopy}. The step sizes were \(0.008^\circ\) and \(0.04^\circ\), respectively. As a result, each voxel in the illuminated layer is associated with  a distribution in $(\phi,\chi)$ angular space. For each voxel, a local orientation was defined by identifying the most intense peak in this  distribution, corresponding to the local lattice orientation of the 
$(02\overline{2})$ planes, perpendicular to the tensile axis. As a result a two-component orientation map is generated. This map is conceptually similar to electron backscatter diffraction (EBSD) maps, but lacks access to the third rotation axis.

The CDD simulations refer to pure Ni and a volume of $8 \times 8 \times 8.485\,\mathrm{\mu m^3}$ being tensile deformed up to 3.5\% strain. An initial dislocation density of 10$^{12}$m$^{-3}$ was introduced in the domain of simulation in the form of random loops distributed roughly equally on all slip system. The crystal was loaded in the [001] direction at a rate of 20 s$^{-1}$ to the above-mentioned strain. The elastic distortion, combining both the elastic strain and lattice rotation as its symmetric and anti-symmetric parts was computed and used in the current work.

The DFXM forward model follows the geometrical–optics formulation of Poulsen \emph{et al.} \cite{Poulsen2021Geometrical-opticsMicroscopy}, in the implementation by Henningsson \emph{et al.} \cite{Henningsson2025}. The input to the model is a spatially resolved and voxelated elastic distortion field, $\mathbf{F}^{\rm e} (\vec{x})$. This field was obtained from the CDD simulations using the kinematic decomposition of El-Azab and Po \cite{ElAzabPo2018},
\begin{equation}
    \mathbf{F}^{\rm e} (\vec{x}) = \boldsymbol{\beta}^{\rm e} (\vec{x}) + \mathbf{I},
\end{equation}
where $\boldsymbol{\beta}^{\rm e}$ is the elastic part of the displacement gradient, $\nabla \mathbf{u}$. The forward model computes synthetic DFXM datasets generated on the same $(\phi,\chi)$ grid as the experiment and processed identically. Implementation details are provided in Appendix B and Appendix C. Calculation of the total dislocation density from the CDD output can be found in Appendix E.

The experimental results are summarised in Figure~\ref{fig:mosaicity_maps}. This shows the evolution of the average lattice orientation across a region-of-interest at four strain levels: 0.2\%, 3.0\%, 5.0\% and 7.6\%. The corresponding experimental stress strain curve is presented in Figure D.2 (a), with the strain levels at which the DFXM measurements were acquired shown in red.

At the initial strain level of 0.2\%, the microstructure is close to homogeneous, consistent with an undeformed single crystal. At 3.0\% strain, the orientation variation increases, but no well-defined dislocation patterning is observed. This behaviour contrasts sharply with [111]-oriented aluminium, where pronounced cellular structures are already established by comparable strain levels \cite{Zelenika2025ObservingDeformation}.

At 5.0\% strain, the most prominent features are planar boundary features that extend across the entire field of view. The spacing is approximately 20 $\mu$m. As mentioned, the emergence of these planar boundaries is not an expected feature for this high-symmetry deformation geometry. In this case, only the faintest indications of dislocation cells are visible. Only at 7.6\% strain does a clear dislocation cell structure emerge, consistent with the microstructures reported in the literature for high-SFE FCC metals. (At this stage, the analysis focuses on a smaller region of interest imaged using a $\times$10 objective, rather than the $\times$2 objective used at lower strains, which accounts for the reduced total lattice rotation observed across the field of view.) Notably, even after the dislocation cell structure has formed, remnants of the earlier planar boundary features persist, suggesting that their formation precedes and influences the subsequent cellular structure.

\begin{figure*}[h!]
	\centering
	\includegraphics[width=0.99\textwidth]{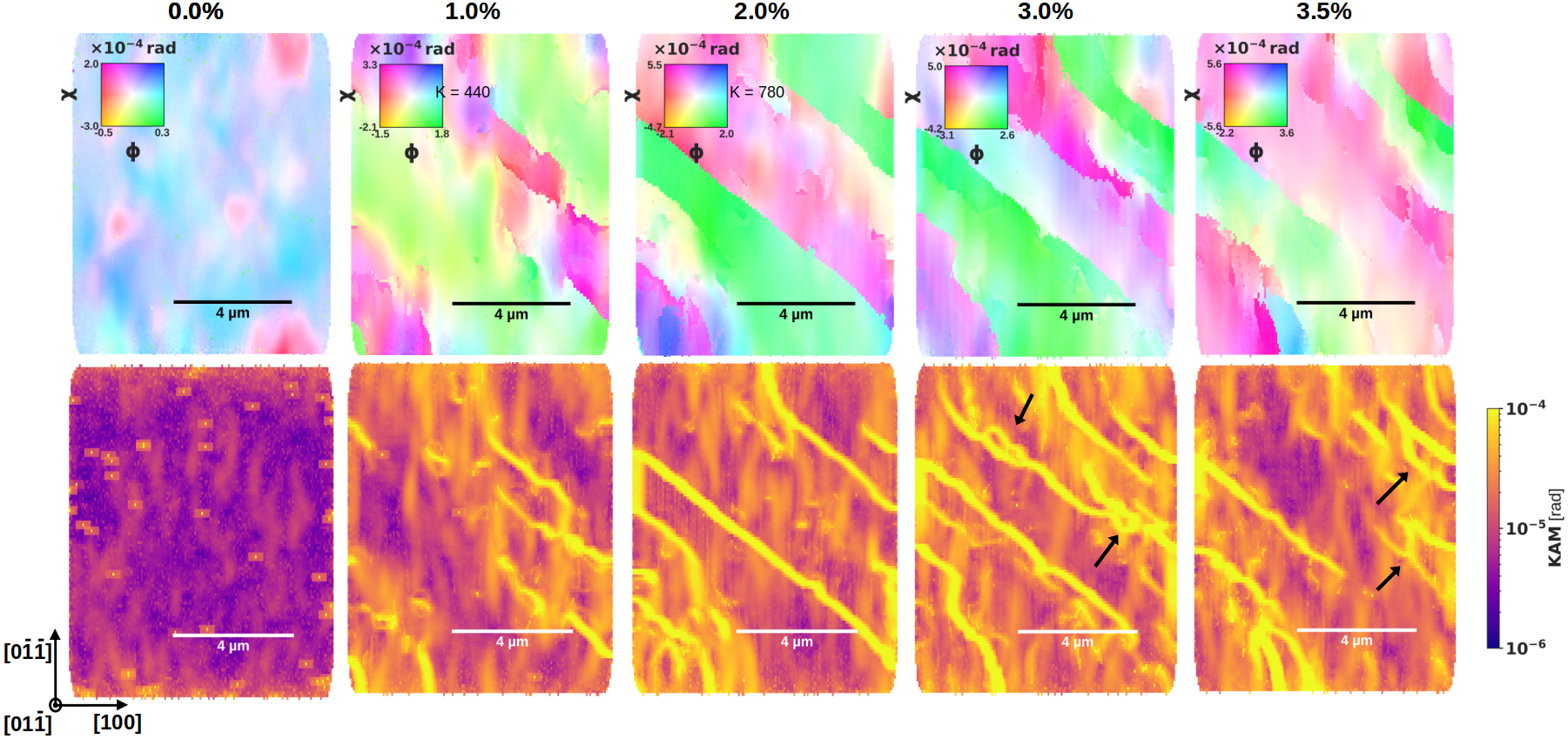}
	\caption{Summary of CDD simulations for [100] tensile deformation of Ni. 
    Top row: Forward-modelled DFXM orientation maps at successive strain levels (0.0\%, 1.0\%, 2.0\%, 3.0\% and 3.5\%). Colour encodes the local lattice orientation. Bottom row: Corresponding kernel average misorientation (KAM) maps (log scale), highlighting local orientation differences, with dislocation cells nucleating highlighted with black arrows.}
	\label{fig:cdd_evolution}
\end{figure*}

The results of the simulations are summarised in Figure~\ref{fig:cdd_evolution}. The upper row shows the evolution of the simulated microstructure obtained from the CDD model, propagated through the DFXM forward model at successive deformation steps. The stress strain curve derived from simulation is presented in Figure D.2 (b), with the analysed deformation states highlighted in red. 

To highlight the boundary formation, we compute kernel average misorientation (KAM) maps (bottom row in Fig.~\ref{fig:cdd_evolution}). KAM, defined as the mean misorientation between a pixel and its nearest neighbours, provides a scalar measure of local orientation gradients. In the simulated sequence it serves as a compact diagnostic for the evolution of the planar boundaries.

Initially, the orientation field in Figure~\ref{fig:cdd_evolution} is seen to be essentially homogeneous, closely resembling an undeformed state of the experimental data. With increasing deformation, diffuse orientation gradients sharpen, leading to the emergence of extended planar boundary features visible in both orientation and KAM maps. The band structure sharpens in the range 1.5 \% to 2.5 \% strain. Analysis of the simulated dislocation density evolution shows that the total dislocation density exhibits a quasi-saturation regime in the same range, as shown in Appendix E. This was interpreted as a temporary stabilisation of the simulated microstructure by SharafEldin \emph{et al.}~\cite{SharafEldin2025}.

At later deformation steps, the planar network progressively loses coherence. The high-KAM bands fragment and the orientation field becomes increasingly heterogeneous, marking the onset of breakup of the planar boundary network and the early nucleation of cells within the strain range simulated in this study.

\begin{figure}[!h]
	\centering
	\includegraphics[width=0.5
    \linewidth]{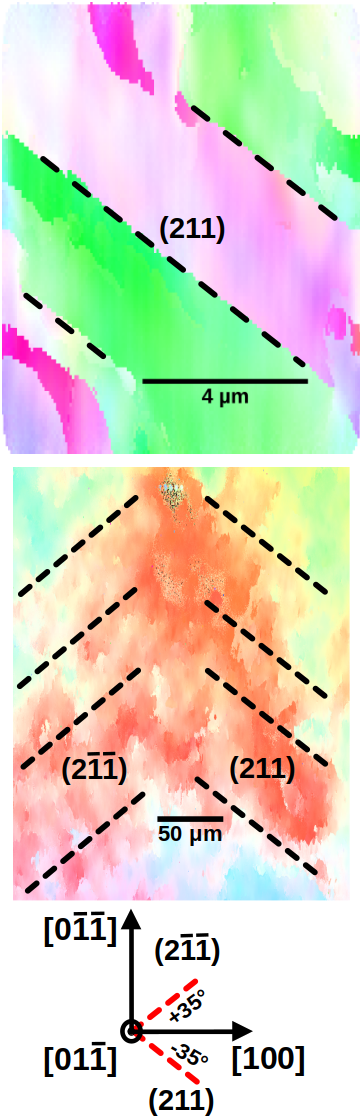}
	\caption{Comparison of planar dislocation boundary trace orientations observed in the CDD simulation (above) and experimentally (below) - for [100]-oriented deformation. In both cases, planar boundaries are visible. The traces align closely with the projected directions of $\{111\}$ slip planes. Dashed-line overlays indicate the calculated slip-trace directions.
	}
	\label{fig:dfxm_trace_overlay}
\end{figure}

In Figure~\ref{fig:dfxm_trace_overlay} we compare the orientations of planar dislocation boundary traces observed experimentally with those obtained from the CDD simulation. In both cases, a single dominant boundary family is observed within the field of view, with trace directions coinciding with the projected traces of $\{111\}$ slip planes for a [100]-oriented crystal. This indicates that the planar boundaries are crystallographically aligned.

A similar pair of boundary orientations was previously reported by Liu and Hansen in cold-rolled cube-oriented aluminium single crystals, where $\{111\}$-aligned planar boundaries were observed at strains of approximately 15\% \cite{Liu1998}. In that study, the boundaries formed within macroscopic transition bands associated with large lattice rotations and shear strains introduced during rolling. In contrast, the present observations demonstrate that similarly aligned planar boundaries emerge at much lower strains, around 5\%, under purely uniaxial tensile loading and within a comparatively homogeneous orientation field.

\begin{figure*}[!h]
	\centering
	\includegraphics[width=0.8
    \textwidth]{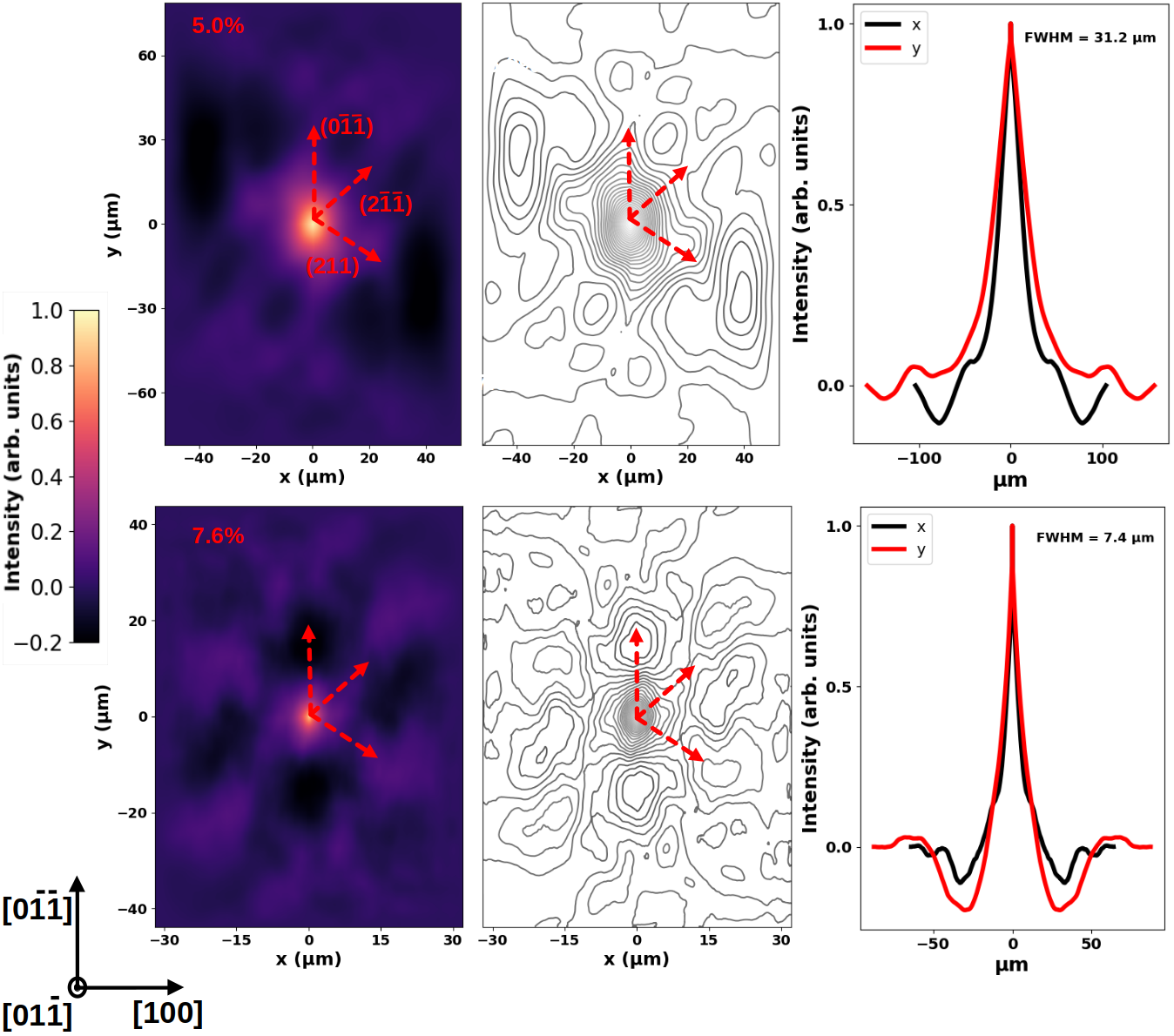}
	\caption{Autocorrelation analysis of experimental rocking-direction orientation maps at 5.0\% strain (above) and 7.6\% strain (below). Red dashed arrows indicate the projected $\{111\}$ slip-trace directions $(0\ \overline{1}\ \overline{1})$, $(2\ 1\ 1)$ and $(2\ \overline{1}\ \overline{1})$. The middle panels show a contour plot highlighting anisotropic correlation features. The right panels show line profiles through the autocorrelation along the horizontal ($x$) and vertical ($y$) directions.}
	\label{fig:autocorr}
\end{figure*}

To quantify whether these planar boundary orientations persist even when individual boundaries are not visually prominent in the orientation maps, we performed an autocorrelation analysis of the experimental maps in the rocking direction, $\phi$. The results are shown in Fig.~\ref{fig:autocorr} for 5.0\% and 7.6\% strain.

At 5.0\% strain, the autocorrelation exhibits an anisotropic response with protrusions aligned with the traces of the $\{111\}$ slip planes, indicated by the red dashed arrows (Note that the traces of $(1\,1\, \overline{1})$ and $(2\,\overline{1}\, 1)$ coincide). In particular, correlation maxima along the $(2\,\overline{1}\, \overline{1})$ direction enable an estimate of the characteristic spacing between planar boundaries of approximately $20~\mu$m, corresponding to the mean separation of successive features projected along this trace direction. The other two trace directions are less sharply defined, but remain discernible in the contour plot in the second panel, indicating that all $\{111\}$ traces are present at this strain level.

The line profiles through the autocorrelation along the horizontal and vertical directions in Fig.~\ref{fig:autocorr} (right) provide a measure of the characteristic coherent domain size, following the interpretation in \cite{Zelenika2025ObservingDeformation}. The full width at half maximum (FWHM) of the central peak yields a length scale of $31.2~\mu$m when estimated via the mean of the $x$ and $y$ FWHM values. The relatively broad central peak and the absence of short-range correlation features associated with a cell network support that, at 5.0\% strain, the microstructure is dominated by comparatively large coherent domains and does not yet exhibit a well-developed dislocation cell structure.

At 7.6\% strain, the autocorrelation changes. The central peak narrows substantially, implying a marked reduction in the coherent domain size and consistent with the formation of a refined dislocation cell structure, as also evident in the local orientation maps in Fig.~\ref{fig:mosaicity_maps}. Using the same combination of the $x$ and $y$ FWHM values yields a characteristic length scale of $7.4~\mu$m. In addition, two pronounced anisotropic features appear in the autocorrelation, aligned with the $(2\,1\,1)$ and $(2\,\overline{1}\, \overline{1})$ trace directions. This demonstrates that the $\{111\}$-aligned planar boundary families remain persistent even after the onset of a clear cellular morphology, despite being less readily identifiable by eye in the orientation maps alone.

The key outcome of this work is that extended planar dislocation boundaries aligned with $\{111\}$ traces form under [100] uniaxial tension before the development of a well-defined dislocation cell structure. The same boundary morphology and crystallographic alignment emerge independently in the CDD simulation when propagated through the DFXM forward model. The agreement is not quantitative: the spacing between the planar boundaries is a factor of five larger in the experimental data, the bands appear earlier in the simulations and the total orientation span differs.  This is expected: the experiment and simulation differ substantially in system size, boundary conditions, effective strain rate and the way dislocations are initially introduced, factors that all are known to influence the evolution. (In our view the difference in sample material,  Al versus Ni, is less important.) However, these differences do not affect the crystallographic selection rules governing boundary orientation. The relevant comparison is therefore geometric and symmetry-based, i.e. whether the same families of boundaries emerge under the same loading configuration.

From this perspective, the correspondence observed here demonstrates that the emergence of $\{111\}$-aligned planar boundaries is an intrinsic consequence of the symmetric multi-slip deformation in [100]-oriented high-SFE FCC crystals. Their appearance in both experiment and simulation indicates that these structures are not artefacts of rolling-induced shear, friction, or macroscopic transition banding, but arise directly from slip system organisation under uniaxial loading. 

The CDD results also show that the persistence of planar boundary networks coincides with a regime in which the total dislocation density exhibits a quasi-saturation, previously identified by SharafEldin \emph{et al.}~\cite{SharafEldin2025} as a temporary stabilisation of the simulated microstructure. In the forward-modelled orientation maps, this regime corresponds to the most coherent and extended planar boundary configurations. At higher deformation steps, as dislocation storage resumes, these planar structures progressively lose coherence and the microstructure begins to fragment, marking the onset of dislocation cell formation. Within the contrast accessible to DFXM and within the strain range simulated here, this evolution mirrors the experimentally observed transition from planar boundary dominated structures to a cellular morphology at larger strains. Finally, the observed progression from an extended planar $\{111\}$ boundary network at intermediate strain to a cell-dominated morphology at higher strain is consistent with a change in the dominant collective organisation of slip with increasing deformation, as discussed in statistical mechanical perspectives on plasticity \cite{Miguel2001,Zaiser2006,Dahmen2009,Friedman2012,Barroso2026}, although no explicit order parameter is identified here.

More broadly, this work establishes a practical framework for linking \emph{in situ} diffraction microscopy to mesoscale dislocation transport models in a common measurement space. By propagating CDD fields through a DFXM forward model and processing simulated and experimental data with identical analysis pipelines, ambiguities associated with representation and post-processing are minimised. This enables robust morphological comparison even when absolute strain scales and rates differ.

Looking forward, an important implication is that DFXM can be applied not only to validate simulations \emph{a posteriori}, but also to initialise CDD simulations from experimentally measured deformed states. Such an approach would allow microstructural evolution to be propagated beyond experimentally accessible strain levels while remaining constrained by experimentally observed dislocation configurations. Realising this requires recovery of the complete elastic deformation gradient field from DFXM data, a goal that recent advances in full-tensor reconstruction are actively bringing within reach \cite{Henningsson2025, Detlefs2025, Kanesalingam2025}. This opens a route toward predictive modelling of dislocation pattern evolution in bulk crystals, grounded in \emph{in situ} experimental input.

\section*{Acknowledgements}
This work was supported by the European Research Council (grant No. 885022), the Danish Agency for Science, Technology, and Innovation through the instrument centre DanScatt (grant No. 7129--00003B), and the Danish Agency for Science and Higher Education (grant No. 8144--00002B). H.F.P. also acknowledges support from a Villum Investigator grant (No. 73771). C.Y. acknowledges financial support from the ERC Starting Grant "D-REX" (grant No. 101116911). A.E and K.S. were supported by the U. S. Department of Energy, Office of Fusion Energy Sciences via award number DE-SC0024585 at Purdue University. We acknowledge ESRF for the provision of synchrotron radiation facilities under proposal No. MA-4442 on beamline ID03.

\bibliography{cas-refs}
\newpage

\clearpage      
\onecolumn

\end{document}